\documentclass[aps,prc,twocolumn]{revtex4}
\usepackage{graphics,amssymb}

\begin{document}

\title{EPR and Bell Locality}
\author{Travis Norsen}
\affiliation{Marlboro College \\ Marlboro, VT  05344 \\ norsen@marlboro.edu}

\date{\today}

\begin{abstract}
A new formulation of the EPR argument is presented, one which uses John
Bell's mathematically precise local causality condition in place of the
looser locality assumption which was used in the original EPR paper and
on which Niels Bohr seems to have based his objection to the EPR argument. 
The new formulation of EPR bears a striking resemblance to Bell's
derivation of his famous inequalities.  The relation between these two 
arguments -- in particular, the role of EPR as part one of Bell's 
two-part argument for nonlocality -- is also discussed in detail.
\end{abstract}

\maketitle

\section{Introduction}
\label{sec:intro}

Eugene P. Wigner summed up a widely held view of
the implications of Bell's Theorem\cite{bell} when he stated:  ``In my 
opinion, the most convincing argument against the theory of hidden variables
was presented by J. S. Bell. ....The...argument shows that any theory 
of hidden variables conforming to the postulate of locality is in 
conflict with quantum mechanics.''\cite{wigner}

N. David Mermin echoed this view in his 
review article on ``Hidden Variables and the Two Theorems of John
Bell'':
\begin{quote}
``Bell's theorem establishes that the value assigned to an observable 
must depend on the complete experimental arrangement under which it 
is measured, even when two arrangements differ only far from the
region in which the value is ascertained -- a fact that Bohm
theory exemplifies, and that is now understood to be an unavoidable
feature of any hidden-variables theory. 

To those for whom nonlocality is anathema, Bell's Theorem finally
spells the death of the hidden-variables program.''\cite{mermin}
\end{quote}
In a nutshell, this common viewpoint seems to be based on the
following apparently straightforward sort of reasoning:  
Bell proved that hidden-variables
theories have to be nonlocal (in order to agree with the empirically 
correct predictions of quantum mechanics); nonlocality conflicts with 
relativity's prohibition on super-luminal causation; relativity is 
true; so hidden-variables theories must be false.  

Yet somehow this argument failed to compell the
discoverer of the theorem in question.  As Mermin concedes, Bell
himself ``did not believe that either of his no-hidden-variables 
theorems excluded the possibility of a deeper level of description 
than quantum mechanics.''
How strange!  \emph{Bell himself did not believe} that what Mermin
refers to as Bell's two ``no-hidden-variables theorems,'' actually 
exclude hidden-variables!  Why didn't Bell accede to the
interpretation of his own theorems offered by Wigner, Mermin, and so 
many other learned commentators on the foundations of quantum
physics?  Was this universally-recognized genius really so obtuse?  

Mermin provides a clue in the continuation of the above block-quote:
\begin{quote}
``But not for Bell.  None of the no-hidden-variables theorems persuaded
him that hidden-variables were impossible.  What Bell's Theorem did
suggest to Bell was the need to reexamine our understanding of Lorentz
invariance...'' 
\end{quote}
Thus Bell believed that his theorems brought out a conflict not merely 
between relativity and hidden-variables theories, but, rather, between 
relativity and the predictions of quantum theory \emph{as such}, in
any interpretation.  Mermin briefly mentions this possible view in a 
footnote:  ``Many people contend that Bell's Theorem demonstrates 
nonlocality independent of a hidden-variables program, but there is 
not general agreement about this.'' 

Evidently the ``many people'' referred to here by Mermin include in
their ranks Bell himself.

Given the unique clarity and forthrightness of Bell's writings, 
it is not surprising that we needn't undertake extensive detective
work to infer Bell's views.  He tells us quite explicitly both 
\emph{that} and \emph{why} he believes his theorems call into 
question our understanding of fundamental space-time structure, and 
not merely the attempt to supplement quantum mechanics with additional 
variables.  

Here is the \emph{that}:  ``...the nonlocality of quantum mechanics 
cannot be attributed to incompleteness, but is somehow
irreducible.''\cite[pg 244]{bell}
Also:  ``The obvious 
definition of `local causality' does not work in quantum mechanics, 
and this cannot be attributed to the `incompleteness' of that theory.''
\cite[pg 256]{bell}   And:  ``For me then this is the real problem
with quantum theory:  the apparently essential conflict between any
sharp formulation and fundamental relativity.  That is to say, we have
an apparent incompatibility, at the deepest level, between the two
fundamental pillars of contemporary theory...'' \cite[pg 172]{bell}

And here is the \emph{why}:  ``That ordinary quantum mechanics is not 
locally causal was pointed out by Einstein, Podolsky, and Rosen, in
1935.'' \cite[pg 24]{bell}  That is, according to Bell, the reason 
Bell's Theorem spells trouble (in the sense of requiring 
apparently-relativity-violating nonlocality)
for more than just hidden-variables theories is that 
standard quantum mechanics itself (regarded as an already-complete 
description of the world) is nonlocal -- a fact which Bell claims 
was pointed out in the famous 1935 EPR paper.

But this raises another mystery.  Everyone knows that the purpose of 
the EPR paper was not to argue that quantum mechanics (QM) is nonlocal,
but, rather, to argue against the completeness doctrine.  Indeed, the 
EPR argument was crucially premised on the very assumption -- locality
-- that Bell claims EPR disproved:  according to EPR, if
\begin{quote}
(A) the quantum-mechanical predictions for certain 
correlations are correct
\end{quote} 
and if 
\begin{quote}
(B) quantum theory is required to respect the principle 
of locality
\end{quote}
then the claim that 
\begin{quote}
(C) the theory is complete
\end{quote} 
cannot be correct.  Or, taking as indubitable the (experimentally 
well-confirmed) empirical predictions of quantum theory referred to 
in (A), the EPR argument takes the form:
\begin{equation}
\mbox{(B)} \rightarrow \neg \mbox{(C)}. 
\end{equation}
That is, locality implies \emph{in}completeness.  (We use the symbol
$\neg$ to denote negation:  e.g., $\neg \mbox{(X)}$ should be read
``it is not the case that (X)''.)

But this is logically equivalent to the claim that
\begin{equation} 
\mbox{(C)} \rightarrow \neg \mbox{(B)}
\end{equation}
(i.e., completeness implies nonlocality) and also to the claim
\begin{equation}
\neg \mbox{(B)} \; \mbox{or} \; \neg \mbox{(C)}.
\label{notBornotC}
\end{equation}
(i.e., either locality or completeness must fail).  
Einstein himself stated the conclusion of the EPR argument in this
last form:
\begin{quote}
``By this way of looking at the matter, it becomes evident that the
paradox [EPR] forces us to relinquish one of the following two
assertions:\\
(1) the description by means of the $\psi$-function is complete. \\
(2) the real states of spatially separated objects are independent of
each other.''\cite{einstein}
\end{quote}

Evidently, then, this is the basis for Bell's assertion that EPR
showed that ``ordinary quantum mechanics is not locally causal.''  For
if we grant the premise (and it is one that has been insisted on by
advocates of the Copenhagen approach ever since the 1930's)
that QM is \emph{complete}, it follows from the EPR argument that 
ordinary QM itself is \emph{nonlocal}.  
So if the EPR argument is sound -- if it 
is correct that quantum mechanics, if complete, is nonlocal -- then 
Bell's own interpretation of the significance of his theorems (and not 
the more widely-held interpretation put forward by Wigner and Mermin) 
would be validated.  For this would show that, whether or not one wishes
to supplement QM with hidden variables, one must advocate a nonlocal theory 
in order to account for the empirical data.

But is the EPR argument sound?  The standard view in the physics community
has been that Niels Bohr refuted the EPR argument in 1935 by pointing out an
``essential ambiguity'' in the famous EPR criterion of reality:  ``If,
without in any way disturbing a system, we can predict with certainty
(i.e., with probability equal to unity) the value of a physical
quantity, then there exists an element of physical reality
corresponding to this physical quantity.''\cite{epr}

Bohr claimed that ``the wording of the above mentioned 
criterion...contains an ambiguity as regards the meaning of the
expression `without in any way disturbing a system'.''  That is, Bohr
seems to have objected to the formulation of \emph{locality} which
entered into the EPR reality criterion.  In particular, he argued that 
there was a type of non-mechanical disturbance which EPR had neglected 
and that the criterion was therefore inapplicable to the very example
on which they base their argument:  ``Of course there is in a case
like that just considered no question of a mechanical disturbance of 
the system under investigation during the last critical stage of the 
measuring procedure.  But even at this
stage there is essentially the question of \emph{an influence on the
very conditions which define the possible types of predictions
regarding the future behavior of the system.}  Since these conditions
constitute an inherent element of the description of any phenomenon to
which the term `physical reality' can be properly attached, we see
that the argumentation of [EPR] does not justify their conclusion that
[, if local, the] quantum-mechanical description is essentially
incomplete.''\cite{bohr}

Many commentators (including, not surprisingly, Bell) have questioned 
the validity, clarity, and relevance of Bohr's reply.  (See
in particular pages 155-6 of Ref. \cite{bell} for Bell's lucid analysis of 
Bohr's reply to EPR.)  Nevertheless, 
it is true that the exact definition of locality
used as a crucial premise in the EPR argument -- and also the exact
role of that premise in the argument -- are less than crystal clear.
It would be desirable, therefore, if the condition of local causality could 
be clarified, and the EPR argument reformulated in terms of this
clearer concept.  This is the goal of the present paper.

Happily, there is almost no work to do to achieve this goal -- for in
the course of establishing his so-called 
``no-hidden-variables'' theorems, John
Bell introduced an intuitive and mathematically precise definition of
local causality.  So the goal at hand can be achieved simply by
replacing EPR's somewhat vague language about not disturbing a distant 
system with the quantitative requirement of Bell Locality.  We will 
perform this replacement in Section \ref{sec:eprbell} after first, 
in Section \ref{sec:epr}, briefly reviewing the original EPR
argument.  Finally, in Section \ref{sec:disc}, we discuss the 
relation of the re-formulated EPR argument to Bell's Theorem -- in 
particular, the role of the EPR argument in Bell's two-part 
argument for nonlocality.

\section{The EPR Argument}
\label{sec:epr}

Before presenting the updated version, let us briefly recap the 
original EPR argument.  We will use the scenario introduced
by Bohm \cite{bohm1951}
in which different spin components of two spin-1/2 particles
take the place of the position and momentum variables used in the
original EPR paper.  The two versions, however, are identical in terms
of logical structure, so we will refer freely to the original EPR
paper as if they had based the argument on Bohm's example.  Our goal in
this section is simply to lay out the logical structure of the EPR
argument, so that we can provide a recognizably similar structure in
the next section (but with Bell Locality in place of EPR's looser 
locality assumption).  

Consider two spin-1/2 particles which are spatially separated and in
the spin singlet state:
\begin{equation}
\psi_0 = \frac{1}{\sqrt{2}} \left( |\!+\!z\!>_1|\!-\!z\!>_2 \,-\;
|\!-\!z\!>_1|\!+\!z\!>_2 \right)
\label{singlet}
\end{equation}
where $|\!+\!z\!>_1$ is the state in which particle $1$ has spin $+$ along
the z-axis, etc.  Equation \ref{singlet} attributes no definite spin values 
to either of the two particles separately, but it does imply a definite
relation between the spins:  whatever the spin of particle $1$ is 
(measured to be) along the z-axis, the spin of particle $2$ along the
z-axis will be (measured to be) opposite.  

The same singlet state can also be written in other bases, e.g., the 
basis of eigenstates for spin along the x-axis:
\begin{equation}
\psi_0 = \frac{1}{\sqrt{2}} \left( |\!+\!x\!>_1 |\!-\!x\!>_2 \,-\; 
|\!-\!x\!>_1 |\!+\!x\!>_2 \right).
\end{equation}
This rewriting makes manifest the fact that, 
in addition to the perfect (anti-) correlation of (measured) spin
values along the z-axis mentioned above, there is also a 
perfect (anti-) correlation of (measured) spin values along the 
x-axis:  if we measure the spin of particle $1$ along the x-axis and
find the result $+$, we can be certain that the spin of particle
$2$ along the x-axis (if/when measured) will be $-$, and vice 
versa.

This perfect anti-correlation of outcomes when the spins are measured
along parallel axes is the first assumption of the EPR argument.  We will 
henceforth refer to it as ``EPR(A)''. 

The EPR argument then proceeds as follows.

According to the EPR criterion of reality, ``If, without in any way
disturbing a system, we can predict with certainty ... the value of a
physical quantity, then there exists an element of physical reality
corresponding to this physical quantity.''  Let
us simply assume that, since particles $1$ and $2$ are spatially
separated, the act of measuring the spin of particle $1$ along the
z-axis doesn't disturb particle $2$ in any way, and likewise for a
measurement of the spin of particle $1$ along the x-axis.  (This
assumption is obviously motivated by relativity's prohibition on
causal relations between space-like separated events -- a relationship
that the two measurement events in question here can simply be
stipulated to have.)  Let us call this locality assumption ``EPR(B)''.  

EPR now argue:  by measuring the z-axis spin of particle $1$, we can 
infer the z-axis spin of particle $2$ by using EPR(A) -- without, 
evidently, in any way disturbing particle $2$.  There exists, 
therefore, an element of reality corresponding to the z-axis spin
of particle $2$.  Why?  Because after the measurement on 
particle $1$, particle $2$ is known to be in a state with a definite 
value of spin along the z-axis.  This follows from a trivial
application of EPR(A).  But, by EPR(B), the measurement on particle 
$1$ could not have caused particle $2$ to acquire this property, for 
particle $2$ was not disturbed in any way by the measurement on $1$.  
Thus if particle $2$ is known to have this property after the 
measurement, it must evidently have possessed it all along, independent 
of the measurement made on particle $1$.  The measurement on $1$, if/when
performed, permits us to \emph{learn} something about the z-axis spin 
of particle $2$.   But the fact we learn about \emph{exists} (i.e., is 
an element of reality) independent of that measurement.

The same argument obviously goes through for the x-axis spin as well:  
by measuring the x-axis spin of particle $1$, we can determine without 
in any way disturbing particle $2$, the x-axis spin of particle $2$.
There exists, therefore, an element of reality corresponding to this 
property of particle $2$ as well.

The reader might perhaps worry that either one or the other
of these inferences can be validly made in a given experimental
situation, but both cannot be, since we can measure at most one 
of the two relevant properties of particle $1$ (and hence infer using
EPR(A) only one of the two relevant properties of particle $2$).  EPR 
answer this possible worry in their paper:
\begin{quote} 
``One could object to this conclusion on the grounds that our criterion 
of reality is not sufficiently restrictive.  
Indeed, one would not arrive at our conclusion if one insisted that 
two or more physical quantities can be regarded as simultaneous 
elements of reality \emph{only when they can be simultaneously 
measured or predicted}.  On this point of view, since either one or 
the other, but not both simultaneously, of the quantities ... can be 
predicted, they are not simultaneously real.  This makes the reality 
of [the two properties of particle 2] depend upon the process of
measurement carried out on [particle 1], which does not disturb the 
second [particle] in any way.  No reasonable [i.e., local] 
definition of reality could be expected to permit this.''\cite{epr}
\end{quote}
Thus -- although the operators for these
two observables don't commute and therefore
cannot according to quantum mechanics possess simultaneous
definite values -- the x-axis and z-axis spins of particle 2 \emph{do}
possess simultaneous definite values.  There are elements of 
reality corresponding to both quantities.  And that means the
descriptive limitations imposed by QM (expressed 
most pointedly by the Heisenberg Uncertainty Principle) can be 
beaten:  there are more facts out there in the world than can be 
squeezed into the quantum mechanical description.\footnote{Although 
the conclusion of incompleteness in the EPR paper
was framed around the notion of beating the uncertainty principle, this
seems not to have been the argument that Einstein himself favored.
See Ref. \cite{norsen}.}  We thus arrive 
at the negation of ``EPR(C)'' -- the claim that the quantum mechanical 
description of reality can be considered complete.

The EPR argument thus takes the symbolic form:
\begin{equation}
\mbox{EPR(A)} \; \mbox{and} \; \mbox{EPR(B)} \, \rightarrow \, \neg \, 
\mbox{EPR(C)}
\end{equation}
which is logically equivalent to
\begin{equation}
\mbox{EPR(A)} \; \mbox{and} \; \mbox{EPR(C)} \, \rightarrow \, \neg \,
\mbox{EPR(B)}
\end{equation}
and also to
\begin{equation}
\neg \, \mbox{EPR(A)} \; \mbox{or} \; \neg \, \mbox{EPR(B)} \; \mbox{or} \; 
\neg \, \mbox{EPR(C)} 
\end{equation}
which is Einstein's formulation quoted above:  (given the empirically
well-verified quantum mechanical expressions for certain correlations)
the advocate of orthodox quantum theory is 
``forced to relinquish'' either the locality claim [EPR(B)] or the 
completeness claim [EPR(C)].

\section{The EPR-Bell Argument}
\label{sec:eprbell}

Like the EPR argument just considered, the new version of EPR (let
us call it the EPR-Bell argument, since it is modeled after
the reasoning of Bell) begins with Bohm's example of a system
consisting of two spatially separated spin-1/2 particles in the 
spin-singlet state, Equation \ref{singlet}.  

Let us first introduce the analog of EPR(A) for the EPR-Bell argument.
We will use 
here a slightly-expanded set of empirical predictions compared to the
simple perfect anti-correlation used in the original EPR argument.
But, like the assumption EPR(A) above, these predictions will all be 
straightforward, uncontroversial predictions of QM that 
are well-confirmed by experiment.  

First, we introduce the probability for joint outcomes ($A$ 
and $B$) for spin measurements along arbitrary directions 
$\hat{a}$ and $\hat{b}$ on the two particles (respectively)
in the singlet state $\psi_0$ of Equation \ref{singlet}:  
\begin{eqnarray}
P(A\!=\!+,B\!=\!+\,|\,\hat{a},\,\hat{b},\,\psi_0) &=& \frac{1}{2} \mbox{sin}^2(\theta/2) \\
P(A\!=\!+,B\!=\!-\,|\,\hat{a},\,\hat{b},\,\psi_0) &=& \frac{1}{2} \mbox{cos}^2(\theta/2) \\
P(A\!=\!-,B\!=\!+\,|\,\hat{a},\,\hat{b},\,\psi_0) &=& \frac{1}{2} \mbox{cos}^2(\theta/2) \\
P(A\!=\!-,B\!=\!-\,|\,\hat{a},\,\hat{b},\,\psi_0) &=& \frac{1}{2} \mbox{sin}^2(\theta/2).
\end{eqnarray}
where $\theta$ is the angle between $\hat{a}$ and $\hat{b}$.
Call this set of assertions ``EPR-Bell(A1)''.  

We also note the standard quantum expressions for the marginal
probabilities for the outcomes of spin measurements on each particle
individually:
\begin{eqnarray}
\label{onehalf1}
&&P(A\!=\!+\,|\,\hat{a},\,\psi_0) = 1/2 \\
\label{onehalf2}
&&P(A\!=\!-\,|\,\hat{a},\,\psi_0) = 1/2 \\
\label{onehalf3}
&&P(B\!=\!+\,|\,\hat{b},\,\psi_0) = 1/2 \\
\label{onehalf4}
&&P(B\!=\!-\,|\,\hat{b},\,\psi_0) = 1/2.
\end{eqnarray}
This set of expressions -- which we shall refer to as
``EPR-Bell(A2)'' -- simply states that 
with the particles in the state $\psi_0$, we are equally likely to 
get a $+$ or $-$ outcome for any single measurement on a single 
particle, independent of the angles $\hat{a}$ and $\hat{b}$. 

With that set of assumptions on the table, let us proceed with the 
argument.  

Consider now a general expression for the joint probability for the 
two outcomes $A$ and $B$, when the spin values along directions 
$\hat{a}$ and $\hat{b}$, respectively, are measured:  
\begin{equation}
\nonumber
P(A,B\,|\,\hat{a},\hat{b},\lambda).
\end{equation}  
Here $\lambda$ is a complete specification of the physical state of the
particle pair prior to measurement.

Let us introduce now ``EPR-Bell(B)'' -- the requirement of Bell
Locality -- according to which the joint
probability $P(A,B\,|\,\hat{a},\hat{b},\lambda)$ 
should factor into a product of individual probabilities 
for the two spatially separated systems, with each factor containing
conditionalization only on local variables.  (See also Bell's 
lengthier and clearer discussion of this result in the essay ``La 
Nouvelle Cuisine'' in Ref. \cite{bell}.)
\begin{eqnarray}
\nonumber 
P(A,B\,|\,\hat{a},\hat{b},\lambda) &=& P(A\,|\,B,\hat{a},\hat{b},\lambda) 
\times P(B\,|\,\hat{a},\hat{b},\lambda) \\ 
\nonumber
&=& P(A\,|\,B,\hat{a},\lambda) \times P(B\,|\,\hat{b},\lambda) \\ 
&=& P(A\,|\,\hat{a},\lambda) \times P(B\,|\,\hat{b},\lambda). 
\label{eprbellb}
\end{eqnarray}
The equality in the first line is standard conditional probability,
and should be completely uncontroversial.
The move from here to the second line involves an application of 
what Abner Shimony has dubbed ``Parameter Independence'' (PI).
\cite{shimony}
This principle asserts that the probabilities associated with 
particular outcomes for each particle should be independent of 
which property is measured on the distant particle -- e.g., 
$P(B|\hat{a},\hat{b},\lambda) = P(B|\hat{b},\lambda)$.  The 
physical intuition motivating PI is that any such stochastic 
dependence could only be accounted for by a ``spooky nonlocal 
action at a distance'' by which the setting of the distant 
instrument somehow causally influenced the (probability distribution 
of) results of the nearby experiment.  

Finally, the move to the third line utilizes what Shimony calls 
``Outcome Independence'' (OI), according to which the probabilities
associated with particular outcomes for each particle should be
independent of the outcome (+ or -) of the distant experiment.  
Again, this seems to be an aspect of the more general (locality)
requirement that what happens \emph{here} should be independent of what
happens \emph{over there} -- or, more precisely, that the (probability
distribution for) outcomes of a given experiment can be completely
accounted for by facts in the past light-cone of the detection event
-- here, by the pre-measurement, complete, joint state of the  
particles: $\lambda$.  

Bell summarizes the motivation for this mathematical statement 
of locality as follows: 
\begin{quote}
``A theory will be said to be locally causal if the probabilities
attached to values of local beables in a space-time region ... are
unaltered by specification of values of local beables in a space-like
separated region..., when what happens in the backward light cone of
[the first region] is already sufficiently specified, for example by
a full specification of local beables [in that backward light cone].''
\cite[pg 240]{bell}
\end{quote}
Bell specifically argues for the validity of the factorization of 
probabilities (our equation \ref{eprbellb} above) as follows:
``Invoking local causality, and the assumed completeness of
... $\lambda$ ... we declare redundant certain of the conditional
variables in the last expression [our first line above] because
they are at space-like separation from the result in question.''
That is, for example, 
$P(B|\hat{a},\hat{b},\lambda) = P(B|\hat{b},\lambda)$, and so forth.

(It should also be noted that Jon Jarrett \cite{jarrett}
was the first to point out that Bell Locality was entailed
by the conjunction of the two principles PI and OI, though
Jarrett referred to these principles by different names.  Tim
Maudlin, however, has quite reasonably criticized Jarrett's
analysis of Bell's factorization principle.  \cite{maudlin}
As Maudlin points out, Jarrett's parsing
is not unique, so the relevance of the distinction between
PI and OI is called into question.  Moreover, Jarrett 
originally argued that a failure of PI would mean a violation 
of relativity, while a failure of OI would not.  But, as Maudlin 
makes clear, this makes no sense:  if relativity is taken to prohibit 
causal dependency between space-like separated events, violations of 
PI and OI are equally at odds with it.  The reader is urged to consult 
Maudlin's text for a much more extensive and highly
enlightening discussion.)

What concerns us here, however,
is not primarily the validity of Bell's mathematical formulation of
local causality, i.e., whether the sort of dependence prohibited by
the factorization of the joint probability is equivalent to the sort 
of dependence that is supposed to be prohibited by relativity.
Rather, our goal is merely to show that this condition (widely known
to be required in Bell's demonstration that hidden variable theories
respecting the condition cannot agree with experiment) can also be 
used to reformulate the EPR demonstration that the orthodox version
of QM (in which the wave function alone is considered a complete
description of reality) must disagree with experiment if it respects
the locality condition. 

Let us then simply name the Bell Locality condition, Equation
\ref{eprbellb}, as one of the premises -- ``EPR-Bell(B)'' -- and
proceed with that argument.

The third principle relevant to our derivation is the completeness 
assumption.  We have previously introduced the symbol $\lambda$ to 
refer to, in Bell's words, ``a full specification of local beables''
in the backwards light cones of the two detection events.
In typical derivations of Bell's theorem, this 
symbol refers to the wave function and/or whatever
hidden-variables are needed to complete the (in that context, 
assumed incomplete) description provided by the wave
function alone.  But we are not here 
reproducing Bell's theorem.  We are instead aiming to reproduce the
EPR argument, so we will assume with Bohr that \emph{quantum mechanics 
itself is already complete, without the addition of any supplementary
``hidden'' variables}.  We will then demonstrate that there is a contradiction 
implied by the four assembled principles, and hence arrive at the EPR 
conclusion that either locality or completeness -- or, less plausibly, 
one of EPR-Bell(A1) or EPR-Bell(A2) -- must fail.

Thus, let us now formally make the completeness assumption -- i.e., 
replace $\lambda$ with the appropriate quantum mechanical wave function,
$\psi_0$.  Call this replacement ``EPR-Bell(C)''.  

We may now combine EPR-Bell(B) with EPR-Bell(C).  The result is:
\begin{equation}
\label{ABfactored}
P(A,B\,|\,\hat{a},\hat{b},\psi_0) = P(A\,|\,\hat{a},\psi_0) \times
P(B\,|\,\hat{b},\psi_0).
\end{equation}
This leaves us in a position to utilize the expressions in EPR-Bell(A2).  
Plugging in yields:
\begin{equation}
\label{onefourth1}
P(A\!=\!+,B\!=\!+\,|\,\hat{a},\hat{b},\psi_0) = 1/4
\end{equation}
and
\begin{equation}
\label{onefourth2}
P(A\!=\!+,B\!=\!-\,|\,\hat{a},\hat{b},\psi_0) = 1/4
\end{equation}
and
\begin{equation}
\label{onefourth3}
P(A\!=\!-,B\!=\!+\,|\,\hat{a},\hat{b},\psi_0) = 1/4
\end{equation}
and
\begin{equation}
\label{onefourth4}
P(A\!=\!-,B\!=\!-\,|\,\hat{a},\hat{b},\psi_0) = 1/4.
\end{equation}

These expressions have been deduced by straightforwardly combining
EPR-Bell(A2), (B), and (C) -- i.e., the quantum expressions for marginal
probabilities, Bell Locality, and the assumption that the quantum 
mechanical description of physical reality is complete.  And, as
should be obvious, these predictions conflict with the expressions in 
EPR-Bell(A1).  We have thus proved that all four of these principles 
cannot be simultaneously correct.  At least one member of the set 
must be false.

And since EPR-Bell(A1) and EPR-Bell(A2) are both directly supported
by experiment, 
we must evidently reject either EPR-Bell(B) -- (Bell) Locality -- or 
EPR-Bell(C) -- completeness -- on pain of contradiction.
\footnote{Perhaps it is worth noting that one currently fashionable
interpretation of QM -- the many-worlds interpretation (MWI) -- regards
QM as both complete and local, eluding EPR's
Locality-Completeness dilemma by denying EPR-Bell(A1) and
EPR-Bell(A2).  In particular, according to MWI these probabilities
are all \emph{zero} since they are probabilities for the experiments to
have definite, specific outcomes -- something which, according to MWI,
never occurs in these situations.}
This of course matches the conclusion reached by EPR.

\section{Discussion}
\label{sec:disc}

We have shown that it is possible to reformulate the EPR argument by 
using Bell's mathematically precise local causality requirement, and 
that doing so permits the EPR argument to go through as intended by
its authors.  That is, we have shown that orthodox QM (to the extent
that it is consistent with experiment and regarded as providing a 
complete description of physical reality) violates Bell Locality.
In some sense, this is very old news.  It has been known for some time
that orthodox QM violates the condition of Outcome Independence (OI),
and that Bell Locality is equivalent to the conjunction of OI and PI.
Yet the full implications of this do not seem to have been
widely appreciated.

Opponents of the hidden-variables program tend to side
with Bohr in dismissing the EPR argument (considered as an argument
against the completeness doctrine), and simultaneously to regard
Bell's Theorem as a valid proof of the non-viability of
hidden-variables theories.  
The real point of the present paper is to demonstrate the
\emph{inconsistency} of these two views, by highlighting the
similarity between the EPR argument (as reformulated here) and Bell's
Theorem. 

There seems to be no way of rejecting the argument in
Section \ref{sec:eprbell} that does not simultaneously commit one to
rejecting the applicability of Bell's Theorem.  For example, one might
object that the factorizability condition (Equation \ref{eprbellb}) 
fails to capture relativity's prohibition on super-luminal causation.
(See, for example, \cite{finefactor}.)  If granted, this objection
would indeed preclude the need to regard orthodox quantum theory as in
conflict with relativity; but it would also remove the ground from
those who use Bell's Theorem to argue that hidden variable theories
must conflict with relativity (and are thus not viable).

Simply put, the EPR argument (as recast in Section \ref{sec:eprbell})
and the well-known arguments leading to Bell's Theorem are
are completely parallel:  each shows that a certain
type of theory, if required to respect the Bell Locality condition,
fails to reproduce certain empirical facts -- or, equivalently, each
shows that the only way the theory in question can maintain
consistency with experiment is to violate Bell Locality.  

In the case of the (reformulated) EPR argument, the relevant theory is the
orthodox interpretation of quantum mechanics, according to which the wave 
function alone is regarded as providing 
a complete description of physical reality.  We may
thus state the upshot of the argument as follows:  if you maintain that
QM is complete (and that its empirical predictions are correct) you
are forced to concede that the theory violates Bell Locality.  Thus,
the completeness assumption entails the failure
of Bell Locality:
\begin{equation}
\mbox{EPR:} \; \mbox{Completeness} \rightarrow \neg \, \mbox{Bell Locality}.
\end{equation}

Bell's Theorem, on the other hand, tells us that a certain type of
local hidden variable theory cannot agree with experiment -- or,
equivalently, the only way a hidden variable theory (i.e., a theory
in which the wave function alone is regarded as an \emph{in}complete
description of physical reality) can be made to agree with experiment 
is to violate the Bell Locality condition:
\begin{equation}
\mbox{Bell:} \; \mbox{Incompleteness} \rightarrow \neg \, \mbox{Bell Locality}.
\end{equation}

Combining these two arguments forces us to conclude (without
qualification, for surely QM either is or is not
complete\footnote{Note that what we are here calling
``Incompleteness'' is, specifically, the assumption that there exist
local random variables which, along with the wave function and the
relevant facts about the detection apparatus, determine the
probability distributions for the possible outcomes of each separate
experiment.  ``Completeness'' is simply the denial of this -- i.e.,
the claim that such variables do not exist.  So it is fully warranted
to assert their disjunction.  Thanks to Arthur Fine (private
communication) for suggesting this clarification.}
)
that Bell Locality fails:
\begin{equation}
\mbox{EPR + Bell}:  \; \neg \, \mbox{Bell Locality}.
\end{equation}

We can now see the extent to which the widely held view described in 
Section \ref{sec:intro} is confused and misleading.  Mermin is, strictly 
speaking, correct when he says:  ``to those for whom
nonlocality is anathema, Bell's Theorem finally spells the death of
the hidden-variables program.''  But he seems to have forgotten that,
to those same people (for whom nonlocality is anathema), the EPR
argument spells the death of the \emph{non-}hidden-variables program 
-- i.e., the orthodox interpretation of QM which upholds the 
completeness doctrine.  For orthodox QM itself violates Bell Locality,
the same locality condition that empirically-viable hidden-variable
theories must, according to Bell's Theorem, violate.

The choice between orthodox QM and hidden 
variables theories is thus not (as is so often suggested
\footnote{It boggles the mind that this can continue to be 
suggested, even though it is abundantly evident that QM itself 
violates OI and hence Bell Locality.}) a choice 
between a local theory and a nonlocal theory; it is a choice
between two non-local theories, two theories 
that violate Bell Locality.  What Bell's 
Theorem (combined with the reformulated EPR argument) spells the death 
of is thus the principle of Bell Locality -- nothing more and nothing less.
People ``for whom [such] nonlocality is anathema'' are therefore simply
out of luck.

This should clarify exactly why Bell understood his theorem not
as ruling out the hidden-variables program, but rather as
evidencing a deep conflict between the predictions of 
quantum theory as such, in any
interpretation, and the locality principle suggested by 
relativity.  This seems to have been misunderstood
largely because it has not been grasped that Bell's Theorem is
the \emph{second part} of a \emph{two-part argument} for the
conclusion.\footnote{The idea and phraseology of a ``two-part argument'' 
is due to Sheldon Goldstein.  I was motivated to write the 
current essay after struggling to come to grips with the arguments 
outlined in Section II of Ref. \cite{goldstein}.  
My goal with the current paper is both to flesh out the
validity of part one of Bell's two-part argument for nonlocality 
-- that is, the validity of EPR -- and also to shed some new light 
on the similarities between the two parts, which I think the 
reformulation of EPR in Section \ref{sec:eprbell} does.  For additional
perspectives on Bell's two-part argument, see also \cite{durretal} and 
\cite{laudisa}.}
The necessary first part of that argument is nothing but EPR, which 
generations of physicists have claimed was refuted by Bohr.  But, 
simply put, it wasn't -- as the new formulation presented in 
Section \ref{sec:eprbell} should help make clear.

In order to stress the parallel structure of the EPR argument and
Bell's Theorem, we have up to this point framed them as both arguments
showing the inevitability of nonlocality for a certain type of
(empirically viable) theory:
EPR shows that the price of regarding QM as complete is rendering
the theory manifestly nonlocal;
Bell shows that if we regard standard QM as
incomplete, the hidden variable theory which replaces it will have
to be nonlocal.  It is possible, however, to frame the same argument
in a slightly different logical form, as follows:  
according to EPR, if we want
to insist that quantum theory respect the Bell Locality principle
(and agree with experiment) we must conclude that it is incomplete
(i.e., that there exist variables which
supplement the wave function description and determine 
experimental outcomes in a local manner); but then, Bell's Theorem
shows that this project cannot succeed -- a theory which uses such
variables to explain the EPR correlations simply cannot 
yield the correct predictions for more general correlations.  
This means that the goal of interpreting quantum mechanics in a 
way that respects the locality principle, cannot be reached.  No
local theory -- orthodox QM most certainly included -- can be
consistent with experiment.  

This is the form in which Bell himself elaborated his complete 
\emph{two-part argument} for nonlocality:
\begin{quote}
``Let me summarize once again the logic that leads to the impasse. 
The EPRB [i.e., EPR-Bohm -- the EPR argument using Bohm's example]
correlations are such that the result of the experiment 
on one side immediately foretells that on the other, whenever the 
analyzers happen to be parallel. If we do not accept the intervention 
on one side as a causal influence on the other, we seem obliged to 
admit that the results on both sides are determined in advance anyway, 
independently of the intervention on the other side, by signals from 
the source and by the local magnet setting. [That is the EPR 
argument -- part 1 of Bell's 2-part argument.]  But this has
implications for non-parallel settings which conflict with those of 
quantum mechanics. [That is Bell's Theorem -- part 2.]  So we 
\emph{cannot} dismiss intervention on one side as a causal influence 
on the other.''\cite[pg 149]{bell}
\end{quote}

It is hoped that the current paper will begin to overturn 
a truly unfortunate historical injustice -- namely, the idea (implied 
by the widely-held misinterpretation of the meaning 
of Bell's Theorem described in Section \ref{sec:intro}) that John Bell 
failed to understand his own most important insight.


\begin{thebibliography}{00}

\bibitem{bell}
John S. Bell, \emph{Speakable
and Unspeakable in Quantum Mechanics}, 2nd ed., 
Cambridge University Press, 2004.

\bibitem{wigner}
Eugene P. Wigner, ``Interpretation of Quantum Mechanics'', 1976, 
reprinted in ``Quantum Theory and Measurement'', John A. Wheeler and 
Wojciech H. Zurek, editors, Princeton University Press, 1983.

\bibitem{mermin}
N. David Mermin, ``Hidden Variables and the Two Theorems of John Bell'',
Rev. Mod. Phys., {\bf{65}}, No. 3, July 1993, pp. 803-815.

\bibitem{einstein}
Albert Einstein, ``Reply to Criticisms'' in
``Albert Einstein: Philosopher Scientist'', P.A. Schilpp, ed.,
Harper and Row, 1959, pg 681.

\bibitem{epr}
Albert Einstein, Boris Podolsky, and Nathan Rosen, ``Can Quantum-Mechanical Description
of Physical Reality be Considered Complete?'', 
Phys. Rev. {\bf{47}}, pp. 777-80 (1935).  Also reprinted in Wheeler and Zurek, 
op cit.

\bibitem{bohr}
Niels Bohr, ``Can Quantum-Mechanical Description
of Physical Reality be Considered Complete?'', Phys. Rev.  {\bf{48}}, 
pp. 696-702; reprinted in Wheeler and Zurek, op cit.  (Emphasis in original)

\bibitem{bohm1951}
David Bohm, ``Quantum Theory'', Prentice-Hall, Inc.,
1951, pages 611-23.

\bibitem{norsen}
Travis Norsen, ``Einstein's Boxes'', Am. J. Phys. {\bf{73}} (2), February
2005, pp. 164-176. 

\bibitem{shimony}
Abner Shimony, ``Our Worldview and Microphysics'' in 
``Philosophical Consequences of Quantum Theory'', James T. Cushing 
and Ernan McMullin, editors, University of Notre Dame Press, 1989

\bibitem{jarrett}
Jon Jarrett, ``On the Physical Significance of the Locality Conditions in
the Bell Arguments'', Nous, {\bf{18}}, pp. 569-89 (1984)

\bibitem{maudlin}
Tim Maudlin, \emph{Quantum Non-Locality and Relativity}, 2nd ed., 
Blackwell Publishing, 2002, Chapter 4.


\bibitem{finefactor}
Arthur Fine, ``Do correlations need to be explained?'', 
in \emph{Philosophical Consequences of Quantum Theory}, edited by
J. Cushing and E. McMullin (University of Notre Dame Press, Notre
Dame, 1989), pp. 175-194; see also Arthur Fine, ``The Shaky Game'',
University of Chicago Press, 1996, pp. 59-63.

\bibitem{goldstein}
Goldstein's article on Bohmian Mechanics at
the Stanford Internet Encyclopedia
(http://plato.stanford.edu/entries/qm-bohm)

\bibitem{durretal}
D. D\"urr, S. Goldstein, and
N. Zanghi, ``Quantum Equilibrium and the Role of Operators as
Observables in Quantum Theory'', Journal of Statistical Physics,
{\bf{116}}, pp. 959-1055 (2004), quant-ph/0308038

\bibitem{laudisa}
F. Laudisa, ``The
EPR Argument in a Relational Interpretation of Quantum Mechanics'',
Foundations of Physics Letters, {\bf{14}} (2), pp. 119-132 (2001).

\end{thebibliography}
\end{document}